\RequirePackage{fix-cm}
\documentclass[smallextended]{svjour3}       
\smartqed  
\usepackage{graphicx}
\usepackage{float}
\usepackage{amsmath}
\usepackage[colorlinks,bookmarks,linkcolor=darkred, citecolor=darkgreen]{hyperref}
\usepackage{color, soul}
\definecolor{darkred}{rgb}{0.9, 0.0, 0.0}
\definecolor{darkgreen}{rgb}{0.0, 0.5, 0.0}

\newcommand{\beq}{\begin{equation}}
\newcommand{\eeq}{\end{equation}}

\newcommand{\ber}{\begin{eqnarray}} 
\newcommand{\eer}{\end{eqnarray}}
\def\bf{\textbf}
\def\slash#1{#1\!\!\!/\!\,\,}

\begin{document}

\title{Radiative corrections to neutron beta decay and (anti)neutrino-nucleon scattering from low-energy effective field theory\thanks{LA-UR-23-20003}
}

\author{Oleksandr Tomalak}


\institute{O. Tomalak \at
              Theoretical Division, T-2, MS B283, LANL\\
              87545 Los Alamos, USA \\
              Tel.: +13312185818\\
              \email{tomalak@lanl.gov} 
}

\date{Received: date / Accepted: date}

\maketitle

\begin{abstract}
We study radiative corrections to neutron beta decay and low-energy (anti)neutrino-nucleon scattering within a top-down effective field theory approach. As it was recently shown, a few electromagnetic and electroweak low-energy coupling constants in heavy-baryon chiral perturbation theory are yet to be determined. Performing matching to the four-fermion effective field theory, we relate these low-energy constants to correlation functions of vector and axial-vector currents. Such relations allow us to explicitly clarify scheme dependence for radiative corrections to neutron decay and low-energy charged-current (anti)neutrino scattering, provide a robust prediction of leading in the electromagnetic coupling constant contributions, and achieve a clear separation between short-distance and long-distance contributions.

\end{abstract}

\vspace{-0.3cm}
\section{Introduction}
\label{intro}

\begin{figure}[tp]
\centering
\includegraphics[width=0.75\textwidth]{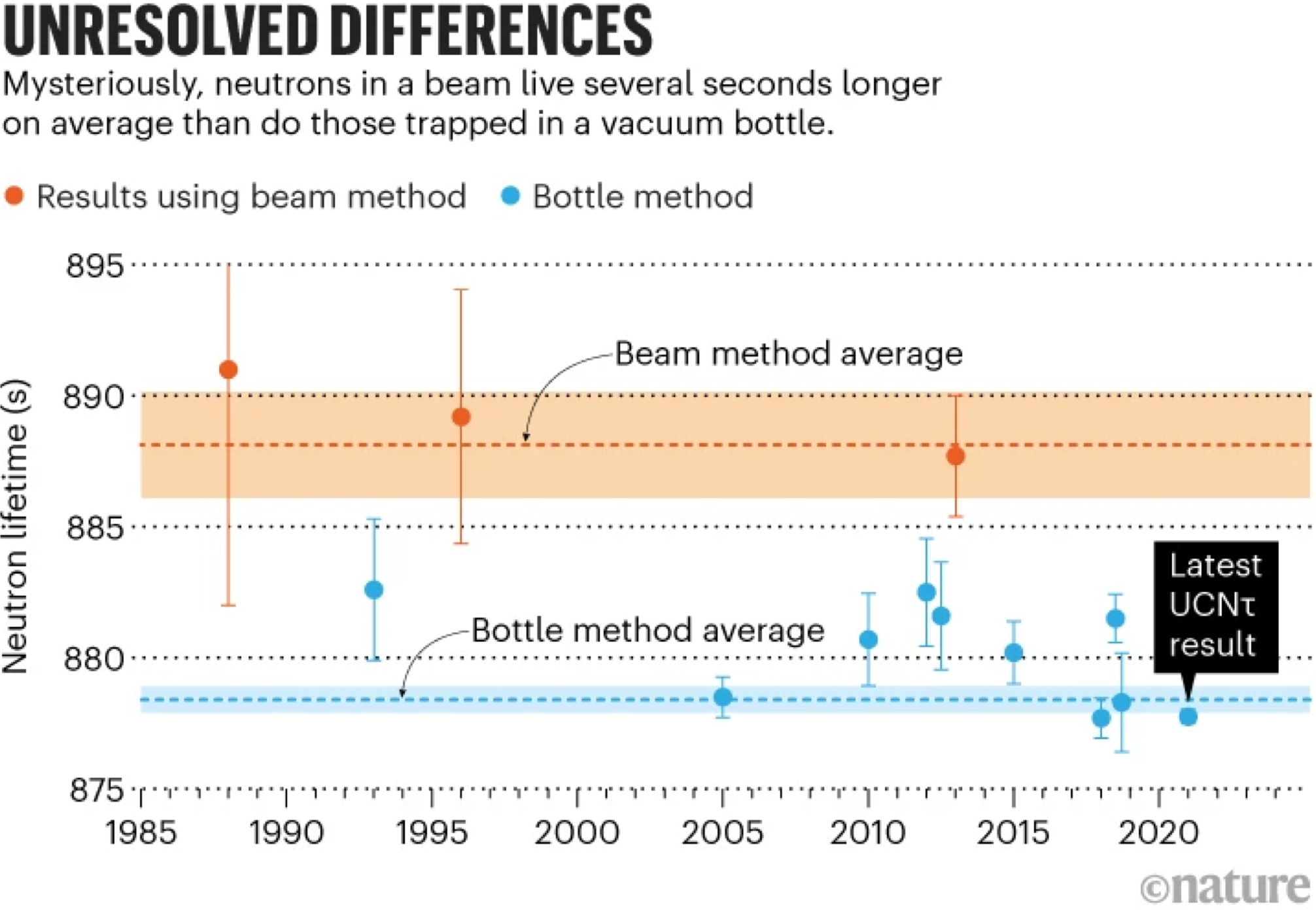}
\caption{Latest measurements of the neutron lifetime by the beam and bottle methods are presented~\cite{Wietfeldt:2018upi,Castelvecchi:2021hec}. \label{fig:neutron_lifetime}}
\end{figure}

Neutron and proton are the lightest baryons. Electroweak interactions trigger the decay of the heavier isospin partner, the neutron, into the proton, the electron, and the electron antineutrino. There are two precise experimental ways to measure the neutron lifetime: beam method and bottle method. The beam method counts the number of slow neutrons that decay while they travel a certain distance. The incoming neutron flux is compared to the number of protons, which are detected in the electric and magnetic field around the neutron decay volume. In the bottle method, the ultracold neutrons are trapped in a reservoir by gravitational or magnetic field. The number of neutrons inside the reservoir is monitored after some period of time. Results that are obtained by these two methods consistently disagree by 8 seconds, which corresponds to 3-to-5 standard deviations~\cite{Wietfeldt:2018upi,Castelvecchi:2021hec}. Such a discrepancy is larger than the uncertainty of the most precise measurements of the neutron lifetime from a single experiment. We illustrate the current status of the neutron lifetime measurements in Fig.~\ref{fig:neutron_lifetime}~\cite{Wietfeldt:2018upi,Castelvecchi:2021hec}.
The latest and most precise measurement by the bottle method was performed at Los Alamos National Laboratory (LANL)~\cite{UCNt:2021pcg}. It reaches a relative  accuracy between 3 and 4 permilles, while theoretical predictions have an error of the same order of magnitude~\cite{Marciano:2005ec,Seng:2018yzq,Seng:2018qru,Czarnecki:2019mwq,Seng:2020wjq,Hayen:2020cxh,Shiells:2020fqp,Hayen:2021iga,Gorchtein:2021fce} and heavily rely on the experimental value of the axial-vector coupling constant. Future measurements by the beam method will decrease the uncertainty by a factor 10; however, we don't know whether that will resolve or deepen the discrepancy.
 
The neutron lifetime measurement in combination with precise extractions of the axial-vector coupling constant is a promising future tool for the determination of the Cabibbo–Kobayashi–Maskawa (CKM) matrix element $V_{ud}$ and tests of the unitarity for this matrix, after resolving the above-mentioned discrepancy. To fully utilize modern and upcoming precise lifetime measurements for the extraction of the physical parameters, an improvement in the theoretical description is required.

\begin{figure}[tp]
\centering
\includegraphics[width=0.6\textwidth]{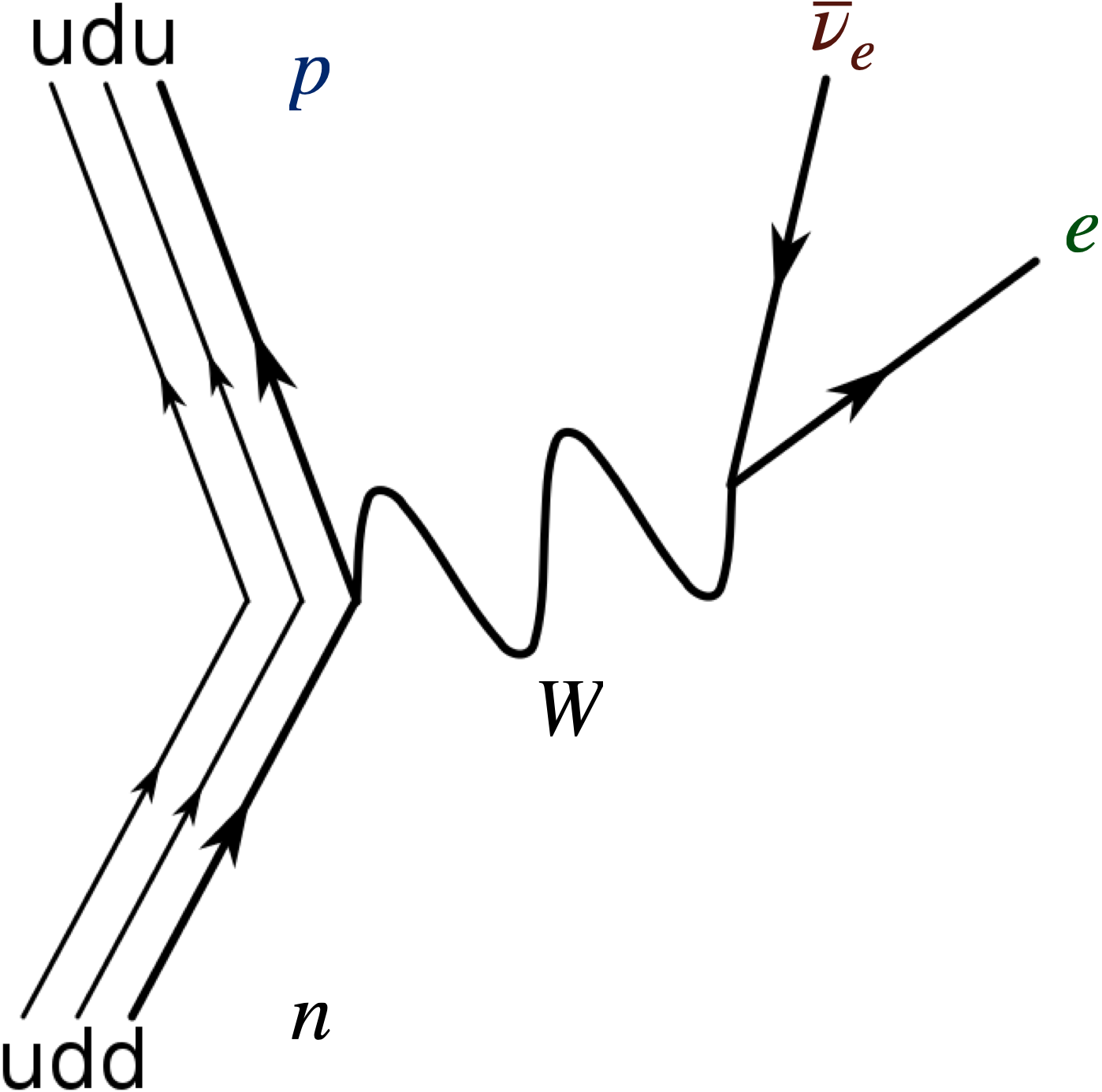}
\caption{The leading-order diagram for neutron decay in the Standard Model. \label{fig:neutron_decay_W}}
\end{figure}

\begin{figure}[tp]
\centering
\includegraphics[width=0.6\textwidth]{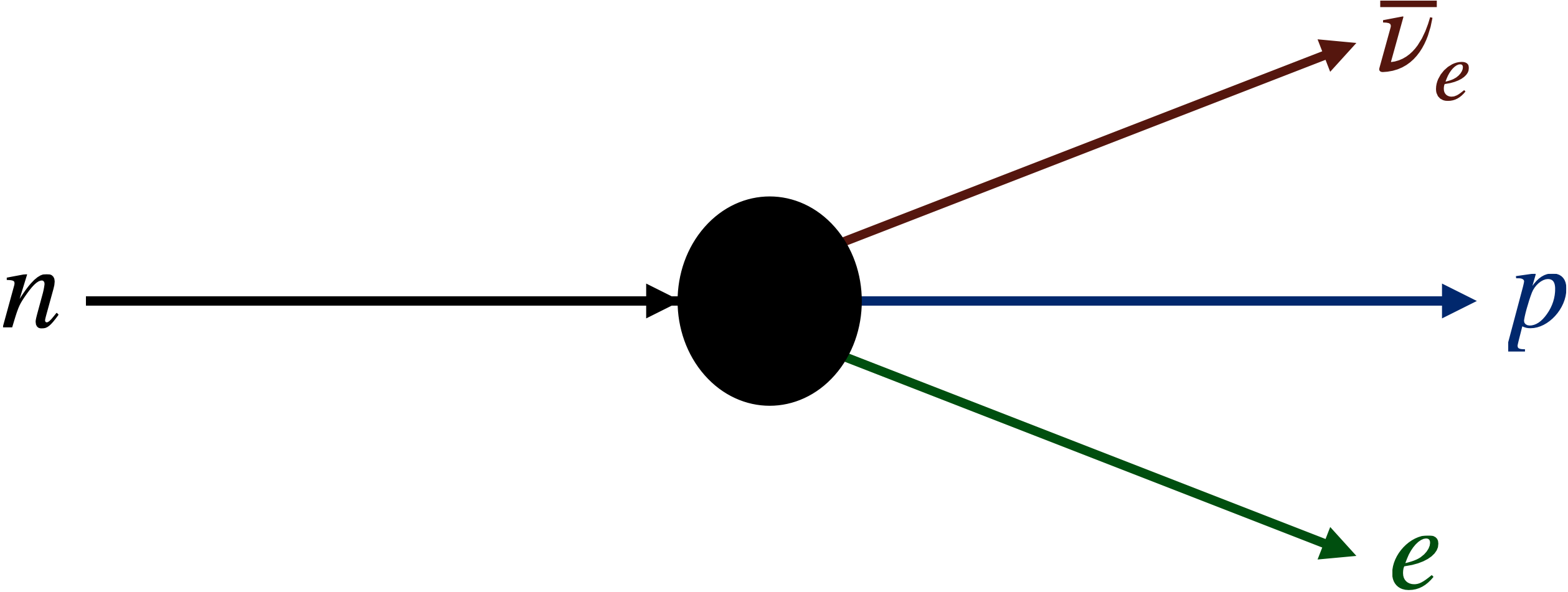}
\caption{The leading-order diagram for neutron decay in the low-energy effective field theory.\label{fig:neutron_decay_EFT}}
\end{figure}

In the Standard Model, the neutron decay is described by the exchange of the electroweak $W$ boson (Fig.~{\ref{fig:neutron_decay_W}}).
However, energy scales for this process ($\sim \mathrm{MeV}$) are well below the electroweak scale ($\sim 100~\mathrm{GeV}$). Consequently, the historical description within the low-energy EFT is the correct physics picture for the neutron decay. We illustrate the leading-order contribution in Fig.~\ref{fig:neutron_decay_EFT}.
The leading-order amplitude for the neutron decay is proportional to the Fermi coupling constant $\mathrm{G}_\mathrm{F}$, which is precisely determined from the muon decay. Charged currents with quarks introduce the CKM matrix element $V_{ud}$ to the neutron beta ($\beta$) decay amplitude. At low energies, the neutron decay is described by two coupling constants: the vector $g_\mathrm{V}$ and the axial-vector $g_\mathrm{A}$. All theoretical works rely on the precise evaluation of electroweak, quantum chromodynamics (QCD), and long-distance hadronic contributions to these coupling constants. In this manuscript, we provide a consistent top-down effective field theory (EFT) description for the neutron decay, starting with the Lagrangian of the Standard Model.

The paper is organized as follows. First, we introduce the low-energy EFT framework for quantum electrodynamics (QED) radiative corrections to the neutron decay in Section~\ref{sec1}. Then, we discuss the bottom-up EFT step of the inclusion of pion degrees of freedom~\cite{Cirigliano:2022hob} in Section~\ref{sec2}. We provide main results for the radiative corrections to the neutron decay within the traditional approach in Section~\ref{sec3}. In Section~\ref{sec4}, we describe the four-fermion EFT with all necessary ingredients following the matching to the Standard Model previously performed in Refs.~{\cite{Hill:2019xqk,Dekens:2019ept}}. The main Section~\ref{sec5} of this paper describes the matching from the four-fermion EFT to the heavy-baryon chiral perturbation theory $\mathrm{HB}\chi\mathrm{PT}$. In the following Section~\ref{sec6}, we provide necessary framework for the evaluation of the hadronic two-current correlation functions. We provide results for the vector coupling and two-current contributions to the axial-vector coupling in Section~\ref{sec7}. Section~\ref{sec8} discusses the relation of the two-current correlators to the experimental data. In Appendix, we provide the notations for the main text of this manuscript.

\section{Low-energy effective theory for neutron decay}
\label{sec1}

The neutron decay is low-energy process, with an energy scale of the order of the neutron-proton mass difference ($1.3~\mathrm{MeV}$). The mass of the electron $m \approx 511~\mathrm{keV}$ is in the same energy range. The scale of the nucleon mass $M$ is 1000 times above the energies of the decay products. Consequently, we can safely work in the EFT considering nucleons as massive particles. Within the bottom-up approach to EFTs, we write the most general form for the interaction Lagrangian, with the vector $g_V$ and the axial-vector $g_\mathrm{A}$ coupling constants~\cite{Ando:2004rk,Raha:2011aa}
\begin{align}
\mathrm{{\cal{L}}}^\mathrm{e^2 p}_\mathrm{\pi N e} = - \sqrt{2} \mathrm{G}_\mathrm{F} V_{ud} \bar{e} \gamma_\rho \mathrm{P}_\mathrm{L} \nu_e \bar{N}_v \left(  g_\mathrm{V} v^\rho - 2 g_\mathrm{A} S^\rho \right) \mathrm{T}^+  N_v  + \mathrm{h.c.} + \mathrm{O} \left( \frac{m}{M}, \alpha \right), \label{eq:Lagrangian_at_leading_order}
\end{align}
where $N_v = \left( p, n \right)^\mathrm{T}$ is the heavy-nucleon field doublet with the nucleon four-momentum $p^\rho$ and velocity $v^\rho = \frac{p^\rho}{M}$, the nucleon spin vector $S^\rho = \frac{1}{2} \left( 0,  \vec{\mathrm{T}} \right)$, the Pauli matrices $\mathrm{T}^i$, $\mathrm{T}^+ = \frac{1}{2} \left( \mathrm{T}^1 + i \mathrm{T}^2 \right)$, and $\alpha \approx 1/137$ is the electromagnetic coupling constant. This Lagrangian is the first step in neutron decay studies. It has already appeared in its covariant form  in the seminal work of Sirlin~\cite{Sirlin:1967zza}. The same Lagrangian describes the low-energy (anti)neutrino-nucleon scattering and is a starting point for the evaluation of inverse $\beta$ decay cross sections.

QED radiative corrections were evaluated from this Lagrangian in Ref.~\cite{Ando:2004rk} (for a detailed study of the radiation of photons, see Ref.~\cite{Bernard:2004cm} in detail). Authors have reproduced the universal radiative correction to low-energy decay rates, which is known as the Sirlin function~\cite{Kinoshita:1958ru,Sirlin:1967zza}, as well as corrections to the angular correlation term~\cite{Garcia:1978bq}. Nucleon recoil corrections were evaluated in Refs.~\cite{Wilkinson:1982hu,Ivanov:2020ybx}. Following the bottom-up approach, authors of Ref.~\cite{Ando:2004rk} introduce counterterms to cancel ultraviolet divergences\footnote{The same ultraviolet divergences appear in the static limit of the (anti)neutrino-nucleon scattering~\cite{Raha:2011aa,Fukugita:2004cq,Tomalak:2021hec,Tomalak:2022xup}.} and propose to fix them from the experimental data.

The traditional approach to radiative corrections in the nucleon $\beta$ decay~\cite{Sirlin:1967zza} matches this EFT directly to the Standard Model in the particular renormalization and regularization scheme without specifying them explicitly (see Section~\ref{sec3}).

To precisely determine the low-energy coupling constants in Eq.~(\ref{eq:Lagrangian_at_leading_order}) at $\mathrm{O} \left( \alpha \right)$, we perform two-step matching from the Standard Model to the four-fermion EFT and to the heavy-baryon chiral perturbation theory in this work. Contrary to the well-known determination of the coupling constants in the meson sector by the two-step matching procedure in Ref.~\cite{Descotes-Genon:2005wrq}, we work everywhere in the well-developed $\overline{\mathrm{MS}}$ renormalization and dimensional regularization schemes.

\section{Role of pions}
\label{sec2}

Recently, next in energy degrees of freedom (pions) were included in Ref.~\cite{Cirigliano:2022hob}, where authors have evaluated the electromagnetic corrections including the virtual pions. According to the results of this work, the vector coupling constant $g_V$ does not renormalize by QED loops in $\chi\mathrm{PT}$ and is solely determined by the same counterterm as in Ref.~\cite{Ando:2004rk}. In contrast, the axial-vector coupling constant is changed by a percent-level contribution from the loops on the $\chi\mathrm{PT}$ side. In the $\overline{\mathrm{MS}}$ renormalization scheme, counterterms of the leading-order Lagrangian in Eq.~(\ref{eq:Lagrangian_at_leading_order}) are given by
\begin{align}
g_\mathrm{V} \left(\mu\right) &= \frac{\alpha}{2 \pi} \hat{C}_\mathrm{V}\left(\mu\right) , \\
g_\mathrm{A} \left(\mu\right)&= \frac{\alpha}{2 \pi} Z_\pi \left[ \frac{1 + 3 \left( g^{\left( 0 \right)}_\mathrm{A} \right)^2}{2} \left( \ln \frac{\mu^2}{m^2_\pi}- 1 \right)- \left( g^{\left( 0 \right)}_\mathrm{A} \right)^2 \right]  +  \frac{\alpha}{2 \pi} \hat{C}_\mathrm{A} \left(\mu\right) g^{\left( 0 \right)}_\mathrm{A},
\end{align}
with the pion mass $m_\pi$, the pion wave-function renormalization constant $Z_\pi$, and the axial charge in the chiral limit and without electromagnetic corrections $g^{\left( 0 \right)}_\mathrm{A}$. $\hat{C}_\mathrm{V}$ and $\hat{C}_\mathrm{A}$ contain the short-distance contributions and hadronic corrections. The determination of these quantities is the main subject of this work. Authors of Ref.~\cite{Cirigliano:2022hob} have also presented the next-to-leading order contribution in $\chi\mathrm{PT}$ with nonzero renormalization of $g_\mathrm{A}$. The change in the axial-vector coupling constant is relevant for the comparison of the lattice QCD calculations of the axial-vector coupling constant to the extractions from the experimental data.

\section{Radiative corrections to neutron decay}
\label{sec3}

Traditionally, radiative corrections to the low-energy electroweak processes are formulated and evaluated within the current algebra approach of Ref.~\cite{Sirlin:1977sv}. This approach was applied to the neutron decay in Refs.~\cite{Sirlin:1967zza,Marciano:1985pd}, where authors have separated large short-distance electroweak logarithms from the smaller long-distance hadronic contributions and universal QED corrections at low energies. At leading order in $\alpha$, the vector and axial-vector coupling constants renormalize with the same short-distance contributions as
\begin{align}
g_\mathrm{V} &= 1 + \frac{\alpha}{2\pi} \left( g + \frac{3}{2} \ln \frac{M_W}{M_A} + \frac{1}{2} \ln \frac{M_W}{M_A} + A_g + C_\mathrm{V} \right), \\
\frac{g_\mathrm{A}}{ g^{\left( 0 \right)}_\mathrm{A}} &=1 + \frac{\alpha}{2\pi} \left( g + \frac{3}{2} \ln \frac{M_W}{M_A} + \frac{1}{2} \ln \frac{M_W}{M_A} + A_g + C_\mathrm{A} \right),
\end{align}
where $g$ is properly averaged over the spectrum in the $\beta$ decay Sirlin's function, $A_g$ denotes perturbative QCD corrections, $C_\mathrm{V}$ and $C_\mathrm{A}$ determine the long-distance hadronic contributions, and the separation scale $M_A$ is of order $1~\mathrm{GeV}$ (modern treatments avoid introducing this scale~\cite{Shiells:2020fqp,Gorchtein:2021fce}). Marciano and Sirlin have also resummed the large electroweak logarithms in their approach~\cite{Marciano:1985pd}. The vector contribution $C_\mathrm{V}$ is expressed in terms of the $\gamma W$ box and can be evaluated with quantified uncertainties. Assuming $C_\mathrm{A} \approx C_\mathrm{V}$~\cite{Czarnecki:2004cw}, the uncertainty of the vector contribution determines the resulting error of the theoretical prediction for the $\beta$ decay rate. According to Ref.~\cite{Czarnecki:2004cw}, the relative theory uncertainty is at the level 4-8$\times10^{-4}$, which is 2 times above the experimental precision for the neutron lifetime~\cite{UCNt:2021pcg}, and urgently requires further reduction.

Recently, this dominant hadronic uncertainty was reduced through the inclusion of details of the physics at low virtualities. Authors of Refs.~\cite{Seng:2018yzq,Seng:2018qru} have provided a model for the pion-nucleon contributions and higher-resonance excitations with consistent treatment of the resonant background, and have connected this region to the high-energy description within the Regge phenomenology and vector-meson dominance model. The updated value for the hadronic contributions to the vector coupling $\frac{\alpha}{2 \pi} C_\mathrm{V} = 3.83(11)\times 10^{-3}$ is a few standard deviations above the result of the previous evaluation $\frac{\alpha}{2 \pi} C_\mathrm{V} = 3.26(19)\times 10^{-3}$~\cite{Czarnecki:2004cw}. This discrepancy resulted in three new dispersive calculations of the $\gamma W$ contribution to $g_V$~\cite{Czarnecki:2019mwq,Hayen:2020cxh,Shiells:2020fqp}, which gave results slightly closer to the recent dispersive determination~\cite{Seng:2018yzq,Seng:2018qru,Seng:2020wjq,Gorchtein:2021fce}; for a concise summary, see Ref.~\cite{Cirigliano:2022yyo}. The first lattice-QCD evaluations of the hadronic contributions to the $\gamma W$ box diagram in the pion and neutron decays are also emerging~\cite{Seng:2020jtz,Feng:2020zdc,Ma:2021azh,Yoo:2022lmt}. Two groups have also recently estimated the difference $C_\mathrm{A} - C_\mathrm{V}$ to be consistent with zero within uncertainties~\cite{Gorchtein:2021fce,Hayen:2021iga}. 

In the following Sections, we describe an alternative treatment of the hadronic and short-distance contributions within the EFT framework for the radiative corrections to neutron decay to provide right physical picture and clear separation between different physics regimes.

\section{Four-fermion effective  field theory}
\label{sec4}

The matching to the Standard Model and the determination of the semileptonic coupling constant $c^{u d}$ in the four-fermion EFT in the neutrino sector, with explicit scale and scheme dependencies, are described in Ref.~\cite{Hill:2019xqk}. Extending the Lagrangian of the four-fermion EFT~\cite{Hill:2019xqk,Cirigliano:2022hob} with external vector $v_\rho$ and axial-vector $a_\rho$ sources, vector ${\bf q}_V$, axial ${\bf q}_A$, and weak ${\bf q}_W$ charge spurions, we obtain
\begin{align}
{\cal L}^\mathrm{spurions}_\mathrm{4-Fermi} &= - e \frac{1}{2} A_\rho \left( \overline{q} {\bf q}_V \gamma^\rho q - \overline{q} {\bf q}_A \gamma^\rho \gamma_5 q\right) + \frac{1}{2}  \overline{q} \left( v_\rho \gamma^\rho - a_\rho \gamma^\rho \gamma_5 \right) q \nonumber \\
&+ \frac{c^{u d}}{2\sqrt{2} \mathrm{G}_\mathrm{F} V_{ud}} \left( \overline{e} \gamma_\rho \mathrm{P}_\mathrm{L} \nu_e \overline{q} {\bf q}_W \gamma^\rho \mathrm{P}_\mathrm{L} q + \mathrm{h.c.}\right).\label{eq:Lagrangian_quark_spurion}
\end{align}
The weak spurion contains a prefactor, such that ${\bf q}_W = -2 \sqrt{2} \mathrm{G}_\mathrm{F} V_{ud} \mathrm{T}^+$ at the physical point. For studies of the renormalization, it is convenient to express the $\mathrm{O} \left( \alpha \right)$ counterterm contributions in the four-fermion EFT with explicit charge factors for quarks and the electron
\begin{align}
{\cal L}^\mathrm{CT}_\mathrm{4-Fermi} &= - 2 \sqrt{2} \mathrm{G}_\mathrm{F} V_{ud} e^2  \left( \overline{e} \gamma_\rho \mathrm{P}_\mathrm{L} \nu_e \overline{d} \gamma^\rho \mathrm{P}_\mathrm{L} u + \mathrm{h.c.}\right) \nonumber \\
&\times \left( g_{00} Q^2_{e} + g_{23} \left( Q_u - Q_d \right)^2 + g_{03} Q_e Q_u + g_{02} Q_e Q_d \right).\label{eq:Lagrangian_four_fermion_gs}
\end{align}

Following Ref.~\cite{Descotes-Genon:2005wrq}, we generalize the Lagrangian of Eq.~(\ref{eq:Lagrangian_four_fermion_gs}) with charge spurions:
\begin{align}
{\cal L}^\mathrm{CT}_\mathrm{4-Fermi} = &- 2 Q^2_e g_{00} e^2 \overline{e} \left( i \slash{\partial} - e Q_e \slash{A} -m \right) e \nonumber \\
&+  Q_e \left( \overline{e} \gamma_\rho \mathrm{P}_\mathrm{L} \nu_e \left( g_{02} e^2 \overline{q} {\bf q}_W  {\bf q}_L \gamma^\rho \mathrm{P}_\mathrm{L} q + g_{03} e^2 \overline{q} {\bf q}_L  {\bf q}_W \gamma^\rho \mathrm{P}_\mathrm{L} q \right) + \mathrm{h.c.} \right) \nonumber \\
&- i g_{23} e^2 \left( \overline{q}_L \left[ {\bf q}_L, D^\rho {\bf q}_L \right] \gamma_\rho q_L + \overline{q}_R \left[ {\bf q}_R, D^\rho {\bf q}_R \right] \gamma_\rho q_R \right), \label{eq:Lagrangian_four_fermion_counterterm}
\end{align}
where $D^\rho {\bf q}_L \equiv \partial^\rho {\bf q}_L - i \left[ l^\rho, {\bf q}_L \right] $ and $D^\rho {\bf q}_R \equiv \partial^\rho {\bf q}_R - i \left[ r^\rho, {\bf q}_R \right] $ are covariant derivatives, and ultraviolet (UV) divergent contributions to the effective couplings~\cite{Descotes-Genon:2005wrq} are given by
\begin{align}
g_{00}^\varepsilon &= - \frac{1}{32\pi^2} \frac{1}{\varepsilon_\mathrm{UV}}, \nonumber \\
g_{23}^\varepsilon &= - \frac{1}{32\pi^2} \frac{1}{\varepsilon_\mathrm{UV}}, \nonumber \\
g_{02}^\varepsilon &= \frac{1}{16\pi^2} \frac{1}{\varepsilon_\mathrm{UV}} , \nonumber \\
g_{03}^\varepsilon &=  -\frac{1}{4\pi^2} \frac{1}{\varepsilon_\mathrm{UV}}, 
\end{align}
where the UV pole $\frac{1}{\varepsilon_\mathrm{UV}}$ in the $\overline{\mathrm{MS}}$ renormalization scheme is related to the space-time dimension $d = 4 - 2 \varepsilon $ as
\begin{align}
\frac{1}{\varepsilon_\mathrm{UV}} &= \frac{1}{\varepsilon} - \gamma_E + \ln \left( 4 \pi\right).
\end{align}
The scale- and scheme-dependent finite one-loop contributions to effective couplings are given by~\cite{Hill:2019xqk,Dekens:2019ept}
\begin{align}
g_{00}^r &= - \frac{1}{32\pi^2} \ln \frac{\mu^2}{\lambda^2} +  \frac{1}{64\pi^2}, \nonumber \\
g_{23}^r &= - \frac{1}{32\pi^2} \ln \frac{\mu^2}{\lambda^2} +  \frac{1}{64\pi^2}, \nonumber \\
g_{02}^r &=  \frac{1}{16\pi^2} \ln \frac{\mu^2}{\lambda^2} + \frac{3}{32\pi^2}  + a\frac{1}{8\pi^2}, \nonumber \\
g_{03}^r &=  -\frac{1}{4\pi^2} \ln \frac{\mu^2}{\lambda^2} + a \frac{1}{8\pi^2} - \frac{3}{16 \pi^2},
\end{align}
where we have performed the calculation in the naive dimensional regularization (NDR) scheme for $\gamma_5$, and the parameter $a$ is defined from the following contraction in $d$ dimensions~\cite{Buras:1989xd,Dugan:1990df,Herrlich:1994kh}:
\begin{align}
 \gamma^\kappa \gamma^\rho \gamma^\sigma \mathrm{P}_\mathrm{L} \otimes \gamma_\sigma \gamma_\rho \gamma_\kappa \mathrm{P}_\mathrm{L} &= 4 \left [ 1 + a \left( 4 - d \right) \right ] \gamma^\rho \mathrm{P}_\mathrm{L} \otimes \gamma_\rho \mathrm{P}_\mathrm{L} + \mathrm{E}. \label{eq:evanescent1} 
\end{align}
The same scheme was used for the determination of the four-fermion coupling constant $c^{u d}$~\cite{Hill:2019xqk}.

\section{Matching of four-fermion theory to heavy-baryon $\chi\mathrm{PT}$}
\label{sec5}

In this Section, we relate low-energy coupling constants to correlation functions in the four-fermion EFT. To derive such relations, we perform the matching by equating the matrix elements of functional derivatives from the generating functional $w$ w.r.t. the spurion fields between $\chi\mathrm{PT}$ and four-fermion calculations, and account for the tree-level and $\mathrm{O} \left( \alpha \right)$ contributions. Schematically,
\begin{align}
&\left( \int \mathrm{d}^d x <  N | \frac{\delta^2 w \left(  {\bf{q}}_W, {\bf{q}}_V, {\bf{q}}_A  \right)}{\delta {\bf{q}}_{j^b} \left( x \right) \delta {\bf{q}}_{i^a} \left( 0 \right)} \Bigg|^\mathrm{tree~level + \mathrm{O} \left( \alpha \right)~loops + \mathrm{ct.}}_{{\bf{q}}=0} | N > \right)_{\chi\mathrm{PT}} = \nonumber \\
&\left( \int \mathrm{d}^d x <  N |\frac{\delta^2 w \left({\bf{q}}_W,   {\bf{q}}_V, {\bf{q}}_A  \right)}{\delta {\bf{q}}_{j^b} \left( x \right) \delta {\bf{q}}_{i^a} \left( 0 \right)} \Bigg|^\mathrm{tree~level + \mathrm{O} \left( \alpha \right)~loops + \mathrm{ct.}}_{{\bf{q}}=0} | N > \right)_{4-\mathrm{Fermi}}, \label{eq:matching}
\end{align}
where $i$ and $j$ stands for vector, axial, and weak, $V,~A$,~and~$W$, respectively, and ct. stands for counterterms. Equation~(\ref{eq:matching}) might be extended with possible momentum insertions and corresponding derivatives, derivatives w.r.t. external sources, and the pion in the initial state. In Sections~\ref{subsec51} and~\ref{subsec52}, we illustrate expressions for the tree-level contributions in $\chi\mathrm{PT}$. In Section~\ref{subsec54}, we provide an explicit form for the counterterm contribution in the 4-Fermi theory. For convenience, we perform matching at the same scale on both $\chi\mathrm{PT}$ and four-fermion sides, which allows us to exactly cancel the gauge dependence and the electron wave-function renormalization.

\subsection{Electromagnetic coupling constants}
\label{subsec51}

Following the derivation in Refs.~\cite{Meissner:1997ii,Gasser:2002am}, we write down the general form for the leading-order electromagnetic Lagrangian in the heavy-baryon sector including all allowed operators and keeping only $<{\cal{Q}}_- > = 0$ constraint
\begin{align}
\mathrm{{\cal{L}}}^\mathrm{e^2 p}_\mathrm{\pi N} &= e^2 \sum \limits_{i = 1}^{14} g_i \bar{N}_v O^{e^2 p}_i N_v, \label{eq:Lagrangian_electromagnetic}
\end{align}
with the explicit form for the operators of interest, which might enter the neutron $\beta$ decay,
\begin{align}
O^{e^2 p}_1 & = <{\cal{Q}}_+^2 -{\cal{Q}}_-^2>  S \cdot u , \qquad  O^{e^2 p}_{11} = i \left[ {\cal{Q}}_+, S \cdot c^- \right], \nonumber \\
O^{e^2 p}_{13} & =  <{\cal{Q}}_+^2 +{\cal{Q}}_-^2> S \cdot u, \qquad O^{e^2 p}_{12} = i \left[ {\cal{Q}}_-, S \cdot c^+ \right], \nonumber \\
O^{e^2 p}_2 & =  <{\cal{Q}}_+ >^2  S \cdot u , \qquad \qquad ~
O^{e^2 p}_9  = \frac{i}{2} \left[ {\cal{Q}}_+, v \cdot c^+ \right]  + \mathrm{h.c.},
\end{align}
where the standard notations are described in Appendix.

Exploiting the spurion technique from Ref.~\cite{Descotes-Genon:2005wrq}, we take derivatives from the generating functional w.r.t. the spurion fields and derive matching relations for the determination of the coupling constants $g_i$. In the following, we provide a few representations for the electromagnetic coupling constants and combinations of interest in terms of correlation functions with and without external pion fields.
\begin{align}
&   \frac{i \varepsilon^{a b c}}{2} v_\rho \frac{\partial}{\partial r_\rho} \left(  \int \mathrm{d}^d x e^{i r \cdot x} <  N | \frac{\delta^2 w \left(  {\bf{q}}_V, {\bf{q}}_A  \right)}{\delta {\bf{q}}_{V^b} \left( x \right) \delta {\bf{q}}_{V^a} \left( 0 \right)} \Bigg|^\mathrm{tree~level}_{{\bf{q}}=0} | N >   \right)  \Bigg|_{r_\rho=0} \nonumber \\
& = e^2 g_9 \bar{N}_v \mathrm{T}^c N_v,  \nonumber\\
&  i \varepsilon^{a b c} \frac{\partial}{\partial r_\rho} \left(  \int \mathrm{d}^d x e^{i r \cdot x} <  N | \frac{\delta^2 w \left(  {\bf{q}}_V, {\bf{q}}_A  \right)}{\delta {\bf{q}}_{V^b} \left( x \right) \delta {\bf{q}}_{A^a} \left( 0 \right)} \Bigg|^\mathrm{tree~level}_{{\bf{q}}=0} | N >   \right)  \Bigg|_{r_\rho=0} \nonumber \\
&= e^2 \left( g_{11} + g_{12} \right) \bar{N}_v S^\rho \mathrm{T}^c N_v, \nonumber \\
 & \frac{1}{5} \left( \delta^{ab} \delta^{cd} - \frac{1}{2} \delta^{ac} \delta^{bd}  \right) \int \mathrm{d}^d y \mathrm{d}^d x <  N | \frac{\delta^3 w \left( v^\mu, a^\mu, {\bf{q}}_V, {\bf{q}}_A \right)}{\delta a^c_\rho \left( y \right)  \delta {\bf{q}}_{V^b} \left( x \right) \delta {\bf{q}}_{V^a} \left( 0 \right)} \Bigg|^\mathrm{tree~level}_{{\bf{q}}=0} | N > \nonumber \\
 &+ \frac{\delta^{c d}}{2} \int \mathrm{d}^d y \mathrm{d}^d x <  N | \frac{\delta^3 w \left( v^\mu, a^\mu, {\bf{q}}_V, {\bf{q}}_A \right)}{\delta a^c_\rho \left( y \right)  \delta {\bf{q}}_{V^0} \left( x \right) \delta {\bf{q}}_{V^0} \left( 0 \right)}\Bigg|^\mathrm{tree~level}_{{\bf{q}}=0} | N > \nonumber \\
 &= e^2 \left(  g_{1} + g_{2} + \frac{g_{11}}{2} + g_{13} \right) \bar{N}_v  \mathrm{T}^d S^\rho N_v + ...,  \qquad a,b,c \neq 0,\nonumber \\
 & -\delta^{a c} \int \mathrm{d}^d y \mathrm{d}^d x <  N | \frac{\delta^3 w \left( v^\mu, a^\mu, {\bf{q}}_V, {\bf{q}}_A \right)}{\delta v^c_\rho \left( y \right)  \delta {\bf{q}}_{A^b} \left( x \right) \delta {\bf{q}}_{V^a} \left( 0 \right)}  \Bigg|^\mathrm{tree~level}_{{\bf{q}}=0} | N > \nonumber \\
 &= e^2 g_{11} \bar{N}_v \mathrm{T}^b S^\rho N_v + ..., \qquad a,b,c \neq 0, \nonumber\\ 
&  \delta^{a b} \int \mathrm{d}^d y \mathrm{d}^d x <  N | \frac{ \delta^3 w \left( v^\mu, a^\mu, {\bf{q}}_V, {\bf{q}}_A \right)}{\delta v^c_\rho \left( y \right)  \delta {\bf{q}}_{A^b} \left( x \right) \delta {\bf{q}}_{V^a} \left( 0 \right)}  \Bigg|^\mathrm{tree~level}_{{\bf{q}}=0} | N > \nonumber \\
&= e^2 \left( g_{11} + g_{12} \right) \bar{N}_v \mathrm{T}^c S^\rho N_v + ..., \qquad a,b,c \neq 0. \label{eq:electromagnetic_LECs}
\end{align}
To cancel the gauge dependence for electromagnetic coupling constants on the level of the matching between $\chi\mathrm{PT}$ and correlation functions of quark currents in four-fermion EFT (at least for the vector coupling), we include also one-loop corrections on the $\chi\mathrm{PT}$ side.  Note that the coupling constant $g_{13}$ always enters the matching expressions and observables in the same combination with the coupling constant $g_1$. Thus, omitting the operator $O^{e^2 p}_{13}$ does not introduce biases in any physical quantity.

\subsection{Electroweak coupling constants}
\label{subsec52}

Reducing the number of assumptions for coordinate-dependent spurions compared to Ref.~\cite{Cirigliano:2022hob}, but still keeping $<{\cal{Q}}^W_L> = 0$ constraint, we obtain the most general weak-interaction Lagrangian in the heavy-baryon sector for the $\mathrm{SU} \left(2 \right)$ isospin group
\begin{align}\mathrm{{\cal{L}}}^\mathrm{e^2 p}_\mathrm{\pi N e} = e^2 \sum \limits_{i=1}^{6} \bar{e} \gamma_\rho \mathrm{P}_\mathrm{L} \nu_e \bar{N}_v \left(  {V}_i v^\rho - 2 {A}_i g^{(0)}_A S^\rho \right) {{\mathrm{O}_i }}  N_v  + \mathrm{h.c.} \label{eq:Lagrangian_electroweak}
\end{align}
We present $\mathrm{{\cal{L}}}^\mathrm{e^2 p}_\mathrm{\pi N e}$ in the same form as the leading-order Lagrangian in Eq.~(\ref{eq:Lagrangian_at_leading_order}), where the operators in Eq.~(\ref{eq:Lagrangian_electroweak}) are given by
\begin{align}
{ \mathrm{O}_1 } &= [{{\cal{Q}}}_{L}, {\cal{Q}}^W_L], \qquad ~~
{ \mathrm{O}_2 } = [{{\cal{Q}}}_{R}, {\cal{Q}}^W_L], \nonumber \\
{ \mathrm{O}_3 } &= \{ {{\cal{Q}}}_{L}, {\cal{Q}}^W_L \}, \qquad ~
{ \mathrm{O}_4 } = \{ {{\cal{Q}}}_{R}, {\cal{Q}}^W_L \}, \nonumber \\
{ \mathrm{O}_5 } &= < {{\cal{Q}}}_{L} {\cal{Q}}^W_L >, \qquad
{ \mathrm{O}_6 } = < {{\cal{Q}}}_{R} {\cal{Q}}^W_L >. \label{eq:operators_electroweak}
\end{align}
Our low-energy constants are related to the couplings in Ref.~\cite{Cirigliano:2022hob} by $\tilde{X}_1 = V_1 + V_3 + V_4 - A_1 - A_3 - A_4$, $\tilde{X}_2 = -V_2$, $\tilde{X}_3 = 2 A_2 g^{(0)}_A$, $\tilde{X}_4 = V_4 + V_6$, and $\tilde{X}_5 = - 2 \left( A_4 + A_6 \right) g^{(0)}_A$. All coupling constants are dimensionless.

All coupling constants in Eq.~(\ref{eq:Lagrangian_electroweak}) are conveniently determined from the functional derivatives w.r.t. the spurion fields in the same combination in which they enter physical observables as
\begin{align}
&  \int \hspace{-0.5pt} \mathrm{d}^d x <\hspace{-0.5pt} e^- \bar{\nu}_e N | \left( \frac{\delta^2 w  \left( {\bf{q}}_V, {\bf{q}}_A, {\bf{q}}_W \right)}{ \delta {\bf{q}}_{V^0} \left( x \right) \delta {\bf{q}}_{W^a} \left( 0 \right)} + \frac{\varepsilon^{a b c}}{2 i}  \frac{\delta^2 w  \left( {\bf{q}}_V, {\bf{q}}_A, {\bf{q}}_W \right)}{ \delta {\bf{q}}_{V^b} \left( x \right) \delta {\bf{q}}_{W^c} \left( 0 \right)}  \right) \Bigg|^\mathrm{tree~level}_{{\bf{q}}=0} | N >   \nonumber \\
&= e^2 \sum \limits_{i=1}^{4} \bar{e}  \gamma_\rho \mathrm{P}_\mathrm{L} \nu_e \bar{N}_v \left(  {V}_i {v}^\rho - 2 {A}_i g^{(0)}_A S^\rho \right)  \mathrm{T}^a N_v. \label{eq:electroweak_LECs}
\end{align}

\subsection{Contribution to beta decay and (anti)neutrino scattering}
\label{subsec53}

Electromagnetic and electroweak low-energy coupling constants of $\chi\mathrm{PT}$ effective Lagrangians in Eqs.~(\ref{eq:Lagrangian_electromagnetic}) and~(\ref{eq:Lagrangian_electroweak}) contribute to the neutron decay and (anti)neutrino-proton scattering through the modification of the vector and axial-vector coupling constants $\hat{C}_\mathrm{V}$ and $\hat{C}_\mathrm{A}$ of the leading-order Lagrangian in Eq.~(\ref{eq:Lagrangian_at_leading_order}) as
\begin{align}
\hat{C}_\mathrm{V} &= 8 \pi^2 \left[ - \frac{X_6}{2} + 2 \left({V}_1 +{V}_2+{V}_3 + {V}_4 \right) - g_9 \right],\label{eq:vector_contribution_LECs} \\
\hat{C}_\mathrm{A} &= 8 \pi^2 \left[ - \frac{X_6}{2} + 2 \left( {A}_1 + {A}_2 + {A}_3 + {A}_4 \right) + \frac{g_1 + g_2 + \frac{g_{11}}{2} + g_{13}}{g_\mathrm{A}^{(0)}} \right], \label{eq:axial_contribution_LECs}
\end{align}
where we have also added the contribution of the pure-leptonic operator and corresponding coupling constant $X_6$ according to Ref.~\cite{Cirigliano:2022hob}. Equations (\ref{eq:vector_contribution_LECs}) and (\ref{eq:axial_contribution_LECs}) generalize Ref.~\cite{Cirigliano:2022hob} by including contributions of the electroweak constants $V_2,~V_3,~V_4,~A_3,~A_4$ and electromagnetic coupling constant $g_{13}$.

\subsection{Renormalization in four-fermion effective field theory}
\label{subsec54}

The evaluation of coupling constants in Eqs.~(\ref{eq:vector_contribution_LECs}) and~(\ref{eq:axial_contribution_LECs}) from the correlation functions in the four-fermion EFT contains divergent integrals in $d=4$. Taking functional derivatives w.r.t. to the spurion fields from the counterterm Lagrangian of the four-fermion EFT in Eq.~(\ref{eq:Lagrangian_four_fermion_counterterm}), we compute corresponding contributions to the vector $\hat{C}^\mathrm{ct.}_\mathrm{V}$ and axial-vector $\hat{C}^\mathrm{ct.}_\mathrm{A}$ coupling constants:
\begin{align}
\hat{C}^\mathrm{ct.}_\mathrm{V} & = \hat{C}^\mathrm{ct.}_\mathrm{A} = 8 \pi^2 \left(- g_{00}^\varepsilon - g_{23}^\varepsilon + \frac{2 g_{03}^\varepsilon - g_{02}^\varepsilon}{3} \right), \label{eq:counterterm_vector_axial} 
\end{align}
where $g_{00}$ is a purely leptonic counterterm. As it is expected from the consistency of the renormalization, the counterterm contribution takes the form of the physical amplitude $g_{00} Q^2_{e} + g_{23} \left( Q_u - Q_d \right)^2 + g_{03} Q_e Q_u + g_{02} Q_e Q_d$, cf. the Lagrangian in Eq.~(\ref{eq:Lagrangian_four_fermion_gs}).

\section{Hadron tensor and operator product expansion}
\label{sec6}

In this paper, we define the hadron tensor ${T}^{\sigma \rho}_{V^b V(A)^a} \left( L,p\right)$ in terms of the two-point correlation function of quark currents, following~\cite{Abers:1968zz}, as:
\begin{align}
   {T}^{\sigma \rho}_{V^b V(A)^a} \left( L, p\right) &= \frac{i}{2} \int \mathrm{d}^d x   e^{i L \cdot x}  <  N  | \mathrm{T} \left[ \overline{q} \mathrm{T}^b \gamma^\sigma q \left( x \right) \overline{q} \mathrm{T}^a \gamma^\rho \left( \gamma_5 \right) q  (0) \right] \hspace{-2pt} | N  > \nonumber \\
   &= \frac{1}{M} \bar{N}_v M^{\sigma \rho} N_v. \label{eq:hadron_tensor}
\end{align}
For applications in $d$ dimensions, we generalize the tensor decomposition of Eq.~(\ref{eq:hadron_tensor})~\cite{Ji:1993ey,Blumlein:1996tp,Maul:1996dx,Blumlein:2012bf}, which is commonly used in deep inelastic scattering literature, by following ideas of Refs.~\cite{Drell:1966kk,Bjorken:1966jh}
\begin{align}
M^{\sigma \rho} & = M^{\sigma \rho}_S + M^{\sigma \rho}_A, \\
M^{\sigma \rho}_S & = \tilde{g}^{\sigma\rho} \left(
\mathrm{T}_1 (\nu, L^2) + \frac{\slash{L} \gamma_5}{M} \mathrm{A}_1 (\nu, L^2) \right) + \frac{\tilde{p}^\sigma \tilde{\gamma}^\rho + \tilde{p}^\rho \tilde{\gamma}^\sigma}{2 M}  \gamma_5 \mathrm{B}_1 (\nu, L^2)
 \nonumber \\
& + \frac{\tilde{p}^\sigma \tilde{p}^\rho}{M^2}  \left(
\mathrm{T}_2 (\nu, L^2) + \frac{\slash{L} \gamma_5}{M} \mathrm{A}_2 (\nu, L^2) \right) + \frac{p^\sigma \tilde{\gamma}^\rho + p^\rho \tilde{\gamma}^\sigma}{2 M}  \gamma_5 \mathrm{B}_2 (\nu, L^2) \nonumber \\
& + \frac{L^{\sigma} L^{\rho}}{M^2}  \left(
\mathrm{T}_4 (\nu, L^2) + \frac{\slash{L} \gamma_5}{M} \mathrm{A}_4 (\nu, L^2) \right)  \nonumber \\
& + \frac{p^\sigma L^{\rho} + p^\rho L^{\sigma}}{2 M^2}  \left(
\mathrm{T}_5 (\nu, L^2) + \frac{\slash{L} \gamma_5}{M} \mathrm{A}_5 (\nu, L^2) \right) , \\
M^{\sigma \rho}_A & =  \frac{1}{2 M} \{ \gamma^{\sigma \rho}, \slash{L} \left( \mathrm{S}_1 (\nu, L^2) + \frac{\gamma_5}{2} \mathrm{T}_3 (\nu, L^2)\right)  + \slash{p} \mathrm{S}_3 (\nu, L^2) \} \nonumber \\
&+ \frac{1}{2M^2} \Big (  [ \gamma^{\sigma}, \gamma^\rho ] L^2 + L^{\sigma} [\gamma^\rho,  \slash{L} ]  + L^{\rho} [ \slash{L},  \gamma^\sigma]  \Big) \mathrm{S}_2 (\nu, L^2),
\end{align}
with $\nu= \frac{p \cdot L}{M}$, $\tilde{g}^{\sigma\rho} = - g^{\sigma\rho} + \frac{L^{\sigma}L^{\rho}}{L^2}$, $\tilde{p}^\sigma = p^{\sigma}-\frac{ \left(p\cdot
L \right)}{L^2}\,L^{\sigma}$, $\gamma^{\sigma \rho} = \frac{1}{2} \left[ \gamma^\sigma, \gamma^\rho \right]$, and $\tilde{\gamma}^\sigma = {\gamma}^\sigma - \frac{L^{\sigma}}{L^2} \slash{L}$. $M^{\sigma \rho}_S$ and $M^{\sigma \rho}_A$ denote the symmetric and antisymmetric contributions, respectively. Note that we have changed the sign in front of the amplitude $\mathrm{T}_3$, as compared to Ref.~\cite{Gorchtein:2021fce}.

The optical theorem relates the imaginary parts of the hadron amplitudes to the proton structure functions as
\ber
 \Im \mathrm{T}_1\left(\nu, L^2\right) & = & \pi F_1\left(\nu, L^2\right), \quad~~~~~~  \Im \mathrm{T}_{2,3,4,5}\left(\nu, L^2\right)  =  \frac{\pi M}{\nu} F_{2,3,4,5}\left(\nu, L^2\right), \nonumber  \\
 \Im \mathrm{S}_{1,3}\left(\nu, L^2\right) & = & \frac{\pi M}{\nu} g_{1,3}\left(\nu, L^2\right), \qquad  ~~ \Im \mathrm{S}_2 \left(\nu, L^2\right)  =  \frac{\pi M^2}{\nu^2} g_2\left(\nu, L^2\right), \nonumber  \\
 \Im \mathrm{S}_3\left(\nu, L^2\right) & = & \frac{\pi M}{\nu} g_3\left(\nu, L^2\right), \qquad ~~~ \Im \mathrm{A}_1 \left(\nu, L^2\right)  =  \frac{\pi M}{\nu} a_1\left(\nu, L^2\right), \nonumber  \\
 \Im \mathrm{A}_{2,4,5}\left(\nu, L^2\right) & = & \frac{\pi M^2}{\nu^2} a_{2,4,5}\left(\nu, L^2\right), ~~ \Im \mathrm{B}_{1,2}\left(\nu, L^2\right) =  \frac{\pi M}{\nu} b_{1,2}\left(\nu, L^2\right).
\eer

Crossing relations for isovector contributions are given by
\begin{align}
F_{1,5} \left( - \nu,~L^2 \right) &= F_{1,5} \left( \nu,~L^2 \right),  \quad 
F_{2,3,4} \left( - \nu,~L^2 \right) = -F_{2,3,4} \left( \nu,~L^2 \right), \nonumber  \\
g_{1,2} \left( - \nu,~L^2 \right) &= - g_{1,2} \left( \nu,~L^2 \right), \qquad
g_{3} \left( - \nu,~L^2 \right) =  g_{3} \left( \nu,~L^2 \right),  \nonumber  \\
a_{1,5} \left( - \nu,~L^2 \right) &= a_{1,5} \left( \nu,~L^2 \right),  \qquad
a_{2,4} \left( - \nu,~L^2 \right) = - a_{2,4} \left( \nu,~L^2 \right),   \nonumber  \\
b_{1,2} \left( - \nu,~L^2 \right) &= - b_{1,2} \left( \nu,~L^2 \right),
\end{align}
while these relations for isoscalar contributions have opposite signs.

The amplitudes $\mathrm{T}_1,~\mathrm{T}_2,~\mathrm{T}_4,~\mathrm{T}_5,~\mathrm{S}_1,~\mathrm{S}_2,~\mathrm{S}_3$ represent vector-vector correlators, while the amplitudes $\mathrm{T}_3,~\mathrm{A}_1,~\mathrm{A}_2,~\mathrm{A}_4,~\mathrm{A}_5,~\mathrm{B}_1,~\mathrm{B}_2$ correspond to vector-axial correlators. For vanishing lepton masses, amplitudes  $\mathrm{T}_4,~\mathrm{T}_5,~\mathrm{A}_4,~\mathrm{A}_5$ do not contribute. The only other tensor structure in Ref.~\cite{Blumlein:2012bf} has a factor of the photon momentum $L$ and, therefore, the corresponding amplitude does not contribute at vanishing lepton masses. Amplitudes $\mathrm{T}_4,~\mathrm{T}_5,~\mathrm{S}_3,~\mathrm{A}_4,~\mathrm{A}_5,~\mathrm{B}_2$ vanish except for nonzero quark masses; consequently, we neglect them in the matching. We are left with the contributions from $\mathrm{T}_1,~\mathrm{T}_2,~\mathrm{T}_3,~\mathrm{S}_1,~\mathrm{S}_2$ as well as $\mathrm{A}_1,~\mathrm{A}_2$, and $\mathrm{B}_1$.

To express the effects of the hadron tensor on the vector and axial-vector coupling constants in terms of the nucleon structure functions, we separate the nucleon pole contributions and write down the dispersion relations accounting for the crossing properties of the invariant amplitudes
\begin{align}
&\mathrm{Re}~\mathrm{T}_3 (\nu, L^2) = \mathrm{Re}~\mathrm{T}_3^\mathrm{pole} (\nu, L^2) +  2 M \int \limits_{\nu^\mathrm{inel}_\mathrm{thr}}^{\infty}\frac{F_3 \left( \nu^\prime,~L^2\right)}{ \nu^{\prime 2}  - \nu^2 } \mathrm{d} \nu^\prime,\nonumber \\
&\mathrm{Re}~\mathrm{S}_1 (\nu, L^2) = \mathrm{Re}~\mathrm{S}_1^\mathrm{pole} (\nu, L^2) + 2 M \int \limits_{\nu^\mathrm{inel}_\mathrm{thr}}^{\infty}\frac{g_1 \left( \nu^\prime,~L^2\right)}{ \nu^{\prime 2}  - \nu^2 } \mathrm{d} \nu^\prime ,\nonumber \\
&\mathrm{Re}~\mathrm{S}_2 (\nu, L^2) =\mathrm{Re}~\mathrm{S}_2^\mathrm{pole} (\nu, L^2) + 2 M^2 \nu\int \limits_{\nu^\mathrm{inel}_\mathrm{thr}}^{\infty}\frac{g_2 \left( \nu^\prime,~L^2\right)}{\nu^{\prime 2} \left( \nu^{\prime 2}  - \nu^2 \right)} \mathrm{d} \nu^\prime, \nonumber \\
&\mathrm{Re} \left( \nu\mathrm{S}_2 (\nu, L^2) \right) =\mathrm{Re} \left( \nu\mathrm{S}_2 (\nu, L^2) \right)^\mathrm{pole} +  2 M^2 \int \limits_{\nu^\mathrm{inel}_\mathrm{thr}}^{\infty}\frac{g_2 \left( \nu^\prime,~L^2\right)}{ \nu^{\prime 2}  - \nu^2} \mathrm{d} \nu^\prime.
\end{align}

To control the UV behavior of hadron-structure contributions to the vector and axial-vector coupling constants, we determine the leading-twist terms of the operator product expansion (OPE) for the nucleon structure amplitudes~\cite{Maul:1996dx,Manohar:1992tz}
\begin{align}
\mathrm{S}_1 (\nu, L^2)&\to \mathrm{S}_3 (\nu, L^2) \to \frac{M^2}{L^2} g^{\left( 0 \right)}_\mathrm{A}, \quad \mathrm{T}_3 (\nu, L^2) \to - 2 \frac{M^2}{L^2}, \quad ~ \mathrm{S}_2(\nu, L^2) <  \frac{1}{L^3}, \nonumber \\
 \mathrm{A}_1 (\nu, L^2) &\to  - \frac{M^2}{L^2} g^{\left( 0 \right)}_\mathrm{A}, \qquad  \qquad 
\quad ~~ \mathrm{B}_1 (\nu, L^2) \to  2 \frac{M}{\nu} g^{\left( 0 \right)}_\mathrm{A}, \quad \mathrm{A}_2(\nu, L^2) <  \frac{1}{L^3}. \label{eq:hadron_amplitudes_OPE}
\end{align}

\section{Vector coupling and two-point correlation functions}
\label{sec7}

Evaluating $\hat{C}_\mathrm{V}$ and $\hat{C}_\mathrm{A}$ from the amplitudes used in Eqs.~(\ref{eq:hadron_amplitudes_OPE}), and adding the counterterm contributions from Eq.~(\ref{eq:counterterm_vector_axial}), we obtain the expected UV poles for the counterterms of the leading-order Lagrangian, cf. Eq.~(\ref{eq:Lagrangian_at_leading_order}):
\begin{align}
\hat{C}_\mathrm{V}^\mathrm{UV} &= \hat{C}_\mathrm{A}^\mathrm{UV} = -\frac{3}{4} \frac{1}{\varepsilon_\mathrm{UV}}. \label{eq:low_energy_counterterms}
\end{align}
With OPE expressions, we also reproduce the expected scheme dependence, which enters through the parameter $a$.

Accounting for the scale and scheme dependence of the coupling constant in the four-fermion EFT, we determine the counterterm for the vector coupling constant of the leading-order Lagrangian, cf. Eq.~(\ref{eq:Lagrangian_at_leading_order}), in the $\overline{\mathrm{MS}}$ renormalization scheme:\footnote{For evaluation of low-energy QED diagrams in the conventional to $\chi\mathrm{PT}$ $\overline{\mathrm{MS}}$ scheme~\cite{Cirigliano:2022hob,Ando:2004rk}, we substitute  $\hat{C}_{\mathrm{V},\mathrm{A}} \to \hat{C}_{\mathrm{V},\mathrm{A}} + \frac{3}{4}$.}
\begin{align}
\hat{C}_\mathrm{V} \left( \mu \right) &=  - \frac{a}{3} + \frac{1}{4} \ln \frac{\mu^2}{\lambda^2} - \ln \frac{\mu^2}{\mu_0^2} + \frac{1}{8} + \frac{8 \pi^2}{e^2} \left(\frac{c^{u d} \left( a, \mu_0 \right)}{2\sqrt{2} \mathrm{G}_\mathrm{F} V_{ud}} - 1 \right) \nonumber \\
&+ \frac{8 \pi^2}{3} \int \frac{i \mathrm{d}^4 L}{\left( 2 \pi \right)^4} \frac{ \nu^2 - L^2}{ L^2 \left( L^2 - \lambda^2 \right)}  \left( \frac{\mathrm{T}_3 (\nu, L^2)}{M^2} + \frac{2}{L^2} \right). \label{eq:vector_isovector_finite_integrals} 
\end{align}
The same counterterm enters the theory with electromagnetic interaction of virtual pions. By this, we provide a well-defined framework for radiative corrections at the nucleon level for the superallowed $0^+ \to 0^+$ Fermi transitions.

An analogous expression for the axial-vector coupling constant is given by
\begin{align}
\hat{C}_\mathrm{A} \left( \mu \right) &= - \frac{a}{3} + \frac{1}{4} \ln \frac{\mu^2}{\lambda^2} -  \ln \frac{\mu^2}{\mu_0^2} +  \frac{1}{8} +  \frac{8 \pi^2}{e^2} \left(\frac{c^{u d} \left( a, \mu_0 \right)}{2\sqrt{2} \mathrm{G}_\mathrm{F} V_{ud}} - 1 \right) \nonumber \\
& + \frac{16 \pi^2}{9}  \int \frac{i \mathrm{d}^4 L}{\left( 2 \pi \right)^4} \frac{ \nu^2 + 2 L^2}{L^2 \left( L^2 - \lambda^2 \right)} \left( \frac{\mathrm{S}_1 (\nu, L^2)}{M^2 g_\mathrm{A}^{(0)}} - \frac{1}{L^2}    \right) \nonumber \\
&+\frac{16 \pi^2}{3} \int \frac{i \mathrm{d}^4 L}{\left( 2 \pi \right)^4} \frac{1}{L^2} \frac{\nu}{M} \frac{\mathrm{S}_2 (\nu, L^2)}{M^2 g_\mathrm{A}^{(0)}}. \label{eq:axial_isovector_finite_integrals}
\end{align}
Eq.~(\ref{eq:axial_isovector_finite_integrals}) properly describes the scale dependence. For correct finite contributions, this expression has to be extended by additional hadron structure contributions from three-current correlation functions, virtual QED, and pion loops on the $\chi\mathrm{PT}$ side.

The isoscalar amplitudes $\mathrm{A}_1,~\mathrm{A}_2$, and $\mathrm{B}_1$ contribute only to the axial-vector coupling constant as
\begin{align}
 \hat{C}^\mathrm{S}_\mathrm{A}  & = \frac{8 \pi^2}{d-1} \int \frac{i\mathrm{d}^d L}{\left( 2 \pi \right)^d} \frac{ \nu^2 - L^2}{ L^2 \left( L^2 - \lambda^2 \right)} \frac{1}{L^2} \frac{\nu}{M} \left( \frac{\mathrm{B}_1 (\nu, L^2)}{g_\mathrm{A}^{(0)}} + \frac{\nu}{M} \frac{\mathrm{A}_2 (\nu, L^2)}{g_\mathrm{A}^{(0)}} \right) \nonumber \\
 &+ 8 \pi^2 \int \frac{i\mathrm{d}^d L}{\left( 2 \pi \right)^d} \frac{ \nu^2 - L^2}{ L^2 \left( L^2 - \lambda^2 \right)} \frac{\mathrm{A}_1 (\nu, L^2)}{ M^2 g_\mathrm{A}^{(0)}}, \label{eq:axial_isoscalar}
\end{align}
with the infrared regulator $\lambda$. However, this contribution can be alternatively expressed in terms of the correlation functions from three-quark currents with a help of Eqs.~(\ref{eq:electromagnetic_LECs}).

At one-loop level, the low-energy coupling constants have an extremely simple form:\footnote{Coincidentally, the same constant term $-11/8$ is obtained in Ref.~\cite{Abers:1968zz}.} 
\begin{align}
\hat{C}_\mathrm{V} \left( \mu \right) &=   \frac{1}{4} \ln \frac{\mu^2}{\lambda^2} - \ln \frac{\mu^2}{M^2_Z} - \frac{11}{8} \nonumber \\
&+ \frac{8 \pi^2}{3} \int \frac{i \mathrm{d}^4 L}{\left( 2 \pi \right)^4} \frac{ \nu^2 - L^2}{ L^2 \left( L^2 - \lambda^2 \right)}  \left( \frac{\mathrm{T}_3 (\nu, L^2)}{M^2} + \frac{2}{L^2} \right), \label{eq:vector_isovector_finite_integrals2}  \\
\hat{C}_\mathrm{A} \left( \mu \right) &=  \frac{1}{4} \ln \frac{\mu^2}{\lambda^2} - \ln \frac{\mu^2}{M^2_Z} -  \frac{11}{8} + \frac{16 \pi^2}{3} \int \frac{i \mathrm{d}^4 L}{\left( 2 \pi \right)^4} \frac{1}{L^2} \frac{\nu}{M} \frac{\mathrm{S}_2 (\nu, L^2)}{M^2 g_\mathrm{A}^{(0)}}  \nonumber \\
&+ \frac{16 \pi^2}{9} \int \frac{i \mathrm{d}^4 L}{\left( 2 \pi \right)^4} \frac{ \nu^2 + 2 L^2}{L^2 \left( L^2 - \lambda^2 \right)}\left( \frac{\mathrm{S}_1 (\nu, L^2)}{M^2 g_\mathrm{A}^{(0)}} - \frac{1}{L^2}  \right).\label{eq:axial_isovector_finite_integrals2}
\end{align}
Compared to the traditional framework, the hadronic contributions enter integrals without electroweak parameters and there is no need to separate $M_A$ from the long-distance contributions. Moreover, all electroweak physics is conveniently included in logarithmic and constant terms. Assuming the validity of the OPE above some scale $\Lambda$, the difference to the traditional approach can be expressed as
\begin{align}
\Delta \hat{C}_\mathrm{V} &= \frac{1}{4} \ln \frac{M_W^2}{M^2_W+\Lambda^2}  + \frac{8 \pi^2}{3} \int \frac{i \mathrm{d}^4 L}{\left( 2 \pi \right)^4} \frac{\nu^2 - L^2}{L^2 \left( L^2 - M^2_W \right)}  \frac{\mathrm{T}_3 (\nu, L^2)}{M^2} \Theta \left(\Lambda^2 + L^2 \right), \label{eq:difference}
\end{align}
with a similar difference for $\hat{C}_\mathrm{A}$, up to the possible contributions discussed around Eq.~(\ref{eq:axial_isoscalar}). As expected, the difference in Eq.~(\ref{eq:difference}) vanishes in the physics case $ M^2 \ll \Lambda^2 \ll M^2_W$ and in the more general limit $\Lambda^2 \ll M^2_W$.

Our resulting expressions in $d=4$ are in agreement with known results~\cite{Marciano:2005ec,Seng:2018yzq,Seng:2018qru,Seng:2020wjq,Hayen:2021iga,Gorchtein:2021fce} in the limit of a large mass of the $W$ boson $M_W \to \infty$. We have also verified our expressions in $d = 4$ by considering the decomposition into the forward scattering amplitudes for the massless electron and in the limits $m \to 0$ and $M \to \infty$ for the amplitudes with the massive electron. Although an extra nucleon spin-flip amplitude is present in the case of massive electrons~\cite{Tomalak:2017owk}, its effects on all observables are suppressed by the electron mass and can be neglected for the phenomenology of the neutron decay.

\section{Relation to the experimental data}
\label{sec8}

Hadron structure in the isospin space is described by isovector amplitudes ${T}_{V^0 V(A)^0}: 1 \otimes 1$, ${T}_{V^0 V(A)^a}: 1 \otimes \mathrm{T}^a$,  ${T}^\mathrm{V}_{V^a V(A)^b}: \mathrm{T}^a \otimes \mathrm{T}^b$ for $a = b$ and isoscalar amplitudes ${T}^\mathrm{S}_{V^a V(A)^b}: \mathrm{T}^a \otimes \mathrm{T}^b$ for $a \neq b$, where we have used the following decomposition for the tensor with two space indices $a, b \neq 0$: ${T}_{V^a V^b} = {T}^\mathrm{V}_{V^a V^b}  \delta^{a b} + i \varepsilon^{a b c} {T}^\mathrm{S}_{V^a V^b} \mathrm{T}^c$. Only ${T}_{V^0 A^a}$ and ${T}_{V^0 V^a},~{T}^\mathrm{S}_{V^a A^b}$ contributions enter the correction to the vector and axial-vector couplings, respectively.

The photon interacts with the quark current through a pure vector contribution with the following isospin structure:
\begin{align}
\frac{1}{6}+ \frac{\mathrm{T}^3}{2}.
\end{align}
Considering the coupling of two photons to the proton and neutron lines separately, we can express all $VV$ contributions in terms of the difference between the proton and neutron amplitudes as
\begin{align}
\left( \frac{1}{36} {T}_{V^0 V^0} + \frac{1}{4} {T}^\mathrm{V}_{V^a V^b} \right) + \frac{1}{6} {T}_{V^0 V^a} \mathrm{T}^a = \frac{{T}^p + {T}^n}{2} + \frac{{T}^p - {T}^n}{2} \mathrm{T}^a,
\end{align}
with ${T}_{V^0 V^a} = 3 \left( {T}^p - {T}^n \right)$. However, such an easy relation does not hold for $VA$ amplitudes. The amplitudes of the type ${T}_{V^0 A^a}$ enter only the correlation function with $\gamma$ and $Z$ couplings, and were constrained by the parity-violating experimental data in Refs.~\cite{Seng:2018yzq} and~\cite{Seng:2018qru}. However, due to the sparsity of the available data, the authors of these works had to rely on modeling for the connection between amplitudes of different isospin. The amplitudes of the type ${T}^\mathrm{S}_{V^a A^b}$ only enter physical processes with the coupling of $\gamma$ and $W$ to the nucleon line.

\section{Conclusions and outlook}
\label{sec9}

We have described the systematic EFT framework for radiative corrections to the neutron $\beta$ decay and low-energy (anti)neutrino-nucleon scattering with a robust separation of short- and long-distance contributions. First, we integrate out electroweak physics and match the Standard Model to the four-fermion EFT. Working in the $\overline{\mathrm{MS}}$ renormalization scheme and dimensional regularization allows us to consistently resum large perturbative logarithms. Performing matching of the four-fermion EFT to the heavy-baryon chiral perturbation theory, we express low-energy coupling constants in terms of correlation functions of quark currents. Compared to the traditional approach, we explicitly specify the scale and scheme dependence in all steps of the calculation. We have finalized the formalism for the determination of the vector low-energy coupling constant and are working on the determination of the axial-vector coupling. Our next steps include an investigation of opportunities for calculating the relevant correlation functions on the lattice, an inclusion of higher-order perturbative contributions, and an evaluation of the long-distance hadronic contributions, based on both the experimental data and the hadronic models.

\appendix{Notations}
\label{app1}

We take the irreducible representation of $\mathrm{SU}(2)$ algebra in the form of the Pauli matrices, i.e., $ [\mathrm{T}^a,\mathrm{T}^b] = 2 i \varepsilon^{a b c} \mathrm{T}^c$, $ \{\mathrm{T}^a,\mathrm{T}^b \} = 2 \delta^{a b}$, and $\mathrm{T}^0 = 1$.

The vector and axial charge spurions are expressed in terms of the left and right charge spurions as
\begin{align}
{\bf{q}}_V &= {\bf{q}}_L + {\bf{q}}_R, \qquad
{\bf{q}}_A = {\bf{q}}_L - {\bf{q}}_R. \label{eq:spurions_VA_vs_LR}
\end{align}
The spurion combinations that enter the gauge-invariant expressions are
\begin{align}
{{\cal{Q}}}^W_L = u {\bf q}_W u^\dagger, \qquad {{\cal{Q}}}_L = u {\bf{q}}_L u^\dagger, \qquad  {{\cal{Q}}}_R = u^\dagger {\bf{q}}_R u, \qquad {{\cal{Q}}}_\pm = \frac{ {{\cal{Q}}}_L \pm {{\cal{Q}}}_R}{2}, \label{eq:spurions_su2}
\end{align}
where the pion fields in the isospin basis $\pi^a$ are combined into the standard fields
\begin{align}
U = u^2 = e^{\frac{i \pi^a \mathrm{T}^a}{F}},\label{eq:pions_su2}
\end{align}
and $F \approx 92.4~\mathrm{MeV}$ is related to the pion decay constant in the chiral limit.

We extend the standard definitions of the $\chi\mathrm{PT}$ building blocks by external sources as
\begin{align}
u_\rho &= i u^\dagger D_\rho U u^\dagger, \quad c_\rho^\pm = - \frac{i}{2} \left( u \left( i \partial_\rho  {\bf{q}}_L +  \left[ l_\rho, {\bf{q}}_L \right]  \right) u^\dagger \pm u^\dagger \left( i \partial_\rho {\bf{q}}_R + \left[ r_\rho,  {\bf{q}}_R \right] \right) u \right), \nonumber \\
u_\rho &= i \{ u^\dagger \left( \partial_\rho - i r_\rho + i e {\bf{q}}_R A_\rho \right) u - u \left( \partial_\rho - i l_\rho - i {\bf q}_W \bar{e} \gamma_\rho \mathrm{P}_\mathrm{L} \nu_e + i e {\bf{q}}_L A_\rho \right) u^\dagger \}, \nonumber \\ \tilde{D}_\rho &= \partial_\rho + \frac{1}{2} \left[ u^\dagger, \partial_\rho u \right] - \frac{i}{2} u^\dagger \left( r_\rho - 
e {\bf{q}}_R A_\rho \right) u - \frac{i}{2} u \left( l_\rho - e {\bf{q}}_L A_\rho - {\bf q}_W  \bar{e} \gamma_\rho \mathrm{P}_\mathrm{L} \nu_e \right) u^\dagger,
\end{align}
without specifying the $\bar{\nu}_e \gamma_\rho \mathrm{P}_\mathrm{L} e$, which is irrelevant for this paper.

We decompose electromagnetic charge spurions for baryons ${\bf{q}}^\mathrm{b}$ and for quarks ${\bf{q}}^\mathrm{q}$ into isoscalar $q^0$ and isovector $q^a$ contributions as
\begin{equation}
{\bf{q}}^\mathrm{b} = q^0 + q^a \mathrm{T}^a, \qquad {\bf{q}}^\mathrm{q} =  \frac{q^0}{3} + q^a \mathrm{T}^a , \label{eq:nucleon_and_quark_spurions}
\end{equation}
with physical values $q_0 = q^a = \frac{1}{2}$ for the left and right spurions, and $q_0 = q^a = 1$ and $0$ for the vector and axial spurions, respectively.

\begin{acknowledgements}
I acknowledge many useful discussions and validations with Emanuele 
Mereghetti, Vincenzo Cirigliano, Richard Hill, and Wouter Dekens, and useful correspondence with Bastian Kubis. I thank Emanuele Mereghetti and Wouter Dekens for reading this manuscript. This work is supported by the US Department of Energy through the Los Alamos National Laboratory and by LANL’s Laboratory Directed Research and Development (LDRD/PRD) program under projects 20210968PRD4 and 20210190ER. Los Alamos National Laboratory is operated by Triad National Security, LLC, for the National Nuclear Security Administration of U.S. Department of Energy (Contract No. 89233218CNA000001). FeynCalc~\cite{Mertig:1990an,Shtabovenko:2016sxi}, LoopTools~\cite{Hahn:1998yk}, and Mathematica~\cite{Mathematica} were extremely useful in this work.
\end{acknowledgements}


\end{document}